\documentclass[12pt]{article}

\def\vector#1{\mbox{\boldmath $#1$}}
\usepackage[dvipdfmx]{graphicx}
\usepackage{subfigure}
\begin{document}

\title{Information cascade, \\
Kirman's ant colony model, \\
 and \\ kinetic Ising model}

\author{Masato Hisakado\footnote{[1]
masato\_hisakadom@yahoo.co.jp}
 \space{} and   Shintaro Mori\footnote{[2] mori@sci.kitasato-u.ac.jp} }

\maketitle

*Standard  \& Poor's, Marunouchi 1-6-5, Chiyoda-ku, Tokyo 100-0005, Japan

\vspace*{1cm}

\dag Department of Physics, School of Science,
Kitasato University, Kitasato 1-15-1, Sagamihara, Kanagawa 252-0373, Japan

 \vspace*{1cm}

\begin{abstract}
In this paper, we discuss a voting model in which voters can obtain information from a finite number of previous voters. There exist three groups of voters:

(i) digital herders and independent voters,

(ii) analog herders and independent voters, and

(iii) $\tanh$-type herders.

In our previous paper \cite{Hisakado3}, we used the mean field approximation for case (i). In that study, if the reference number $r$ is above three, phase transition occurs and the solution converges to one of the equilibria.
However, the conclusion is different from mean field approximation. In this paper, we show that the solution oscillates between the two states.
A good (bad) equilibrium  is  where a majority of $r$  select the correct (wrong) candidate.
In this paper, we show that there is no phase transition when $r$ is finite.
If the annealing schedule is adequately slow from finite $r$ to infinite $r$, the voting rate converges only to the good equilibrium.

In case (ii), the state of reference votes is equivalent to that of Kirman's ant colony model, and it follows beta binomial distribution.

In case (iii), we show that the model is equivalent to the finite-size kinetic Ising model. If the voters are rational, a simple herding experiment of information cascade is conducted. Information cascade results from the quenching of the kinetic Ising model.
As case (i) is the limit of case (iii) when  $\tanh$ function becomes a step function, the phase transition can be observed in infinite size limit. We can confirm that there is no phase transition when the reference number $r$ is finite.
\end{abstract}


\newpage
\section{Introduction}

While collective herding behaviour is popularly studied among animals, it can also be observed in human beings. In this regard, there are interesting problems across the fields of sociology \cite{tarde}, social psychology \cite{mil}, ethnology \cite{fish}\cite{frank}, and economics. Statistical physics has been an effective tool to analyse these macro phenomena among human beings and has led to the development of an associated field, sociophysics \cite{galam}\cite{Cas}. For example, in statistical  physics, anomalous fluctuations in financial markets \cite{Cont}\cite{Egu} and opinion dynamics \cite{Stau}\cite{Curty}\cite{nuno}\cite{Fil1}\cite{Fil2}\cite{Ly}\cite{Ar}\cite{And} have been discussed.

Most individuals observe the actions of other individuals to estimate public perception and then make a choice similar to that of the others; this is called social learning. Because it is usually sensible to do what other people are doing, collective herding behaviour is assumed to be the result of a rational choice that is based on public perception. While this approach could be viable in some ordinary cases, as a macro phenomenon, it can sometimes lead to arbitrary or even erroneous decisions. This phenomenon is known as an information cascade \cite{Bikhchandani}. In this paper, we show that an information cascade is described by the Ising model.

In our previous paper, we introduced a sequential voting model \cite{Hisakado2}. At each time step $t$, one voter opts for either of two candidates. As public perception, the $t$th voter can see all previous votes, that is, $(t-1)$ votes. To identify the relationship between information cascade and phase transition, we introduce two types of voters---herders and independents. We also introduce two candidates.

The herders' behaviour is known as the influence response function, and threshold rules have been derived for a variety of relevant theoretical scenarios representing this function. Some empirical and experimental evidence supports the assumption that individuals follow threshold rules when making decisions in the presence of social influence \cite{watts2}. This rule posits that individuals will switch from one decision to another only when sufficiently many others have adopted the other decision. Such individuals are called digital herders \cite{Hisakado3}. From our experiments, we observed that human beings exhibit a behaviour between that of digital and analog herders, that is, the tanh-type herder \cite{Hisakado4}. We obtained the probability that a herder makes a choice under the influence of his/her prior voters' votes. This probability can be fitted by a $\tanh$ function \cite{Mori3}.


Here, we discuss a voting model with two candidates. We set two types of voters: independents and herders. As their name suggests, the independents collect information independently, that is, their voting depends on their fundamental values and rationality. In contrast, the voting of herders is based on public perception, which is visible to them in the form of previous votes. In this study, we consider the case wherein a voter can see the latest $r$ previous votes.

When $r\rightarrow \infty$ is the upper limit of $t$, we can observe several phenomena \cite{Hisakado3}. In the case where there are independent voters and digital herders,
the independents cause the distribution of votes to converge to one-peak distribution, a Dirac measure when the ratio of herders is small. However, if the ratio of herders increases above the transition point, we can observe the information cascade transition. As the fraction of herders increases, the model features a phase transition beyond which a state where most voters make the correct choice coexists with one where most of them are wrong. Further, the distribution of votes changes from one peak to two peaks.

In the previous paper,  we discussed the finite $r$ case \cite{Hisakado3}. We analysed the model by using mean field approximations and concluded that information cascade transition occurs when $r\geq3$. In this paper, we discuss the model from other perspectives and show that there is no phase transition when $r$  is finite and the solution oscillates between two equilibria.
Furthermore, we show  relations among our voting model, Kirman's ant colony model, and kinetic Ising model.




The remainder of this paper is organized as follows. In section 2, we introduce our voting model and mathematically define the two types of voters---independents and herders. In section 3, we discuss the case where there are digital herders and independents. In section 4, we verify the transitions between voting choices through numerical simulations. In section 5, we discuss the case where there are analog herders and independents and show the relation between our voting model and Kirman's ant colony model. In section 6, we discuss the relation between our voting model and the kinetic Ising model. In the final section, we present the conclusions.

\section{Model}


We model the voting of two candidates, $C_0$ and $C_1$. The voting is sequential, and at time $t$, $C_0$ and $C_1$ have $c_0(t)$ and $c_1(t)$ votes, respectively. In each time step, one voter votes for one candidate. Hence, at time $t$, the $t$th voter votes, after which the total number of votes is $t$. Voters are allowed to see just $r$ previous votes for each candidate; thus, they are aware of public perception. $r$ is a constant number.


We assume an infinite number of two types of voters---independents and herders. The independents vote for $C_0$ and $C_1$ with probabilities $1-q$ and $q$, respectively. Their votes are independent of others' votes, that is, their votes are based on their own fundamental values.

Here, we set $C_0$ as the wrong candidate and $C_1$ as the correct one in order to validate the performance of the herders. We can set $q\geq 0.5$, because we  believe that independents vote for $C_1$ rather than for $C_0$. In other words, we assume that the intelligence of the independents is virtually accurate.

In contrast, the herders' votes are based on the number of previous $r$ votes. At time $t$, the information of $r$ previous votes are the number of votes for $C_0$ and $C_1$:$c_0^r
(t)$ and $c^r_1(t)$, respectively. Hence, $c_0^r(t)+c_1^r(t)=r$ holds. If $r > t$, voters can see $t$ previous votes for each candidate. For the limit $r \rightarrow \infty$, voters can see all previous votes. We define the number of all previous votes for $C_0$ and $C_1$ as $c_0^\infty (t)\equiv c_0(t)$ and $c_1^\infty(t)\equiv c_1(t)$.

Now we define the majority's correct decision.
If the ratio to the candidate $C_1$ who is correct is $c_1 /t >1/2(<1/2)$, we define the majority as correct (wrong).
This ratio is important to evaluate the performance of the herders.
In this paper, we consider three kinds of herders, namely digital, analog, and tanh-type herders.  We define $z_r$ as $z_r= c_1^r/ r$. The probability that a herder  who refers $z_r$ votes to the candidate $C_1$ is defined as $f(z_r)$. Digital herders always choose the candidate with a majority of the previous $r$ votes, which is visible to them \cite{Hisakado3}. In this case, $f(z_r)=\theta(z_r-1/2)$, where $\theta$ is a heaviside function.  Analog herders vote for each candidate with probabilities that are proportional to the candidates' votes \cite{Hisakado2}.  Thus, $f(z_r)=z_r$. The other herder is the tanh-type herder, who is an intermediate between analog and digital herders \cite{Hisakado4}. In this case, $f(z_r)=1/2(\tanh (z_r-1/2)+1)$.



The independents and herders appear randomly and vote. We set the ratio of independents to herders as $(1-p)/p$. In this study, we mainly focus on the upper  limit of $t$. This refers to the voting of infinite voters.

\section{Digital herder case}

In this section, the herder is a digital herder \cite{Hisakado3}, and hence, votes for the majority candidate; if $c^r
_0 (t) >
c^r
_1 (t)$, herders vote for the candidate $C_0$. If  $c^r
_0 (t) <
c^r
_1 (t)$, herders vote for
the candidate $C_1$. If $c^r
_0 (t) = c^r
_1 (t)$, herders vote for $C_0$ and $C_1$ with the
same probability, that is, 1/2.

In the previous paper, we discussed this case by using mean field approximations \cite{Hisakado3}. In this paper, we discuss this using stochastic partial differential equations.
First, we consider an  approximation to introduce the partial differential equations. The voter selects $r$ votes randomly from the latest previous $r$ voters with overlapping. We can write the process as
\begin{eqnarray}
c_1(t)&=&k \rightarrow k+1:
 P_{k,t:l,t-r}=p\pi((k-l)/r)+(1-p)q,
\nonumber \\
c_1(t)&=&k   \rightarrow k:
 Q_{k,t:lt-r}=1-P_{k,t},
\label{pd}
\end{eqnarray}
where $c_1(t-r)=l$.
$P_{k,t}$ and $Q_{k,t}$  are the probabilities of the process.
The sum of $P_{k,t}$ and $Q_{k,t}$ is $1$.
This means that at time $(t-r)$, the number of votes for $C_1$ is $c_1 (t-r)=l$. Here, we define $\pi(Z)$ as the majority  probability of binomial distributions of $Z$.
$\pi(Z)$ can be calculated as follows:
\begin{equation}
\pi(Z)=\frac{(2n+1)!}{(n!)^2}\int_0^{Z}x^n(1-x)^ndx=\frac{1}{B(n+1,n+1)}\int_0^{Z}x^n(1-x)^ndx,
\label{pi}
\end{equation}
where $r=2n+1$.

For convenience, we define a new variable $\Delta_t$ such that
\begin{equation}
\Delta_t=2c_1(t)-t=c_1(t)-c_0(t).
\label{d}
\end{equation}
We change the notation from $k$ and $l$ to $\Delta_t$ and $\Delta_{t-r}$ for convenience. Given $\Delta_t=u$ and $\Delta_{t-r}=s$, we obtain a random walk model:
\begin{eqnarray}
\Delta_t&=&u \rightarrow u+1  :P_{u,t:s,t-r}=\pi(\frac{1}{2}+\frac{u-s}{2r})p+(1-p)q,
\nonumber \\
\Delta_t&=&u \rightarrow u-1  :Q_{u,t:s,t-r}=1-P_{u,t}.
\nonumber
\end{eqnarray}
We now consider the continuous limit $\epsilon \rightarrow 0$,
\begin{eqnarray}
X_{\hat{\tau}}&=&\epsilon\Delta_{[t/\epsilon]},
\nonumber \\
P(x,\hat{\tau})&=&\epsilon P(\Delta_t/\epsilon,t/\epsilon),
\end{eqnarray}
where $\hat{\tau}=t/\epsilon$,$\hat{r}=r/\epsilon$ and $x=\Delta_t/\epsilon$.
On approaching the continuous limit, we can obtain the stochastic partial differential equation (see Appendix A):
\begin{equation}
\textrm{d}X_{\hat{\tau}}=\biggl[(1-p)(2q-1)-p+2p\frac{(2n+1)!}{(n!)^2}\int_0^{\frac{1}{2}+\frac{X_{\hat{\tau}}-X_{\hat{\tau}-\hat{r}}}{2r}}x^n(1-x)^ndx\biggr]\textrm{d}\hat{\tau}+\sqrt{\epsilon},
\label{ito13}
\end{equation}
where we used (\ref{pi}).
The equation (\ref{ito13}) depends on $X_{\hat{\tau}-\hat{r}}$ and is the feedback system.

We are interested in the behaviour of $X_{\hat{\tau}}$ in the limit $\hat{\tau}\rightarrow \infty$.
The relation between $X_{\infty}$ and the voting ratio to $C_1$,$Z$  is $2Z-1=X_{\infty}/\hat{\tau}$.

We can assume the stationary solution to be
\begin{equation}
X_\infty=\bar{v}\hat{\tau}+(1-p)(2q-1)\hat{\tau},
\label{h3}
\end{equation}
where $\bar{v}$ is constant.
Substituting (\ref{h3}) into (\ref{ito13}), we can obtain
\begin{equation}
\bar{v}=-p+\frac{2p\cdot(2n+1)!}{(n!)^2}\int_0^{\frac{1}{2}+\frac{(1-p)(2q-1)}{2}+
\frac{\bar{v}}{2}} x^n(1-x)^ndx.
\label{i3}
\end{equation}
Equation (\ref{i3}) is self-consistent with a permission of overlapping.
It is the same as the  mean field approximation in \cite{Hisakado3}.
If we set $n=2r+1\rightarrow  \infty$, which is the second term of RHS, the herders behave as digital herders and (\ref{i3}) becomes the strict form for them.


When $r=1,2$, there is only one solution of (\ref{i3}) in the range $p$. However, when $r\geq 3$, there is only one solution in $p<p_c$ and three solutions in $p>p_c$, where $p_c$ is critical $p$ \cite{Hisakado3}. The middle solution is unstable, while the other two solutions are stable. As there exist both good and bad equilibria, there may be phase transitions.

In the case where $r\rightarrow \infty$, the description  is correct. In fact, there is phase transition when $r$ is infinite. The voting rates converge to one of the two stable points. However, when $r$ is finite, the solution oscillates between good and bad equilibria. Hence, there is no phase transition. We elaborate this below.

Here, we consider a random walk  between  the two states, $c_1^{r} (t)/r>1/2$ (good equilibrium) and $c_0^{r}(t)/r>1/2$ (bad equilibrium).\footnote{It is coarse-grained votes and  corresponds to the block spin transformation \cite{mig}.} We define the hopping probability from the state $c_0^r (t)/r>1/2$ to $c_1^r (t)/r>1/2$ as $a$ and
that from the state $c_1^{r} (t)/r>1/2$ to $c_0^{r} (t)/r>1/2$ as $b$. Note that $a$ and $b$ are not the function of $t$. When $t>r$, the transition matrix $A$ of this random walk is
\begin{equation}
A=\left (
\begin{array}{ll}
1-a& a \nonumber \\
b & 1-b
\end{array}
\right ).
\label{pro}
\end{equation}
The random walk of the two states, $\tilde{X}_n$, is defined as the transition matrix $A$ when $t>r$ and the initial condition $\tilde{X}_0=0$. If $r > t$, voters can see $t$ previous votes for each candidate.

The model was studied as a one-dimensional correlated random walk \cite{Bohm}\cite{Konno}. For the limit $t\rightarrow \infty$,
\begin{equation}
\tilde{X}_{\infty} \Longrightarrow  N(\frac{a-b}{a+b}t,\frac{4ab(2-(a+b))}{(a+b)^3} t),
\label{theo}
\end{equation}
where $N(\mu,\sigma^2)$ is the normal distribution with mean $\mu$ and variance $\sigma^2$. (See Theorem 3.1, in \cite{Bohm}). $a$ and $b$ are given in (\ref{pro}).

If consecutive independent voters choose the candidate $C_1(C_0)$ when $c_0^{r} (t)/r>1/2 (c_1^{r} (t)/r>1/2) $, the state changes from $c_0 ^{r} (t)/r>1/2 (c_1^{r} (t)/r>1/2) $ to $c_1^{r} (t)/r>1/2 (c_0^r (t)/r>1/2) $. Thus, independent voters behave as a switch for hopping. When the independents voters who vote $C_1(C_0)$ are the  majority, the state hops from $c_0^r (t)/r>1/2 (c_1^r (t)/r>1/2) $ to $c_1^r (t)/r>1/2 (c_0^r (t)/r>1/2)$.
Hence, the hopping rates $a$ and $b$ are estimated to be
\begin{eqnarray}
a&=&\pi[(1-p)q]=\frac{(2n+1)!}{(n!)^2}\int_0^{(1-p)q}x^n(1-x)^ndx\sim (1-p)^{\frac{r+1}{2}}q^{\frac{r+1}{2}}, \nonumber \\
b&=&
\pi[(1-p)(1-q)]
\sim (1-p)^{\frac{r+1}{2}}(1-q)^{\frac{r+1}{2}},
\label{pi1}
\end{eqnarray}
where the approximations are in $p\sim 1$. In the case where $r=1$, $a=(1-p)q$ and $b=(1-p)(1-q)$. We obtained an identical solution in \cite{Hisakado2}.

In the finite $r$ case, the hopping rates $a$ and $b$ do not decrease as $t$ increases and the state oscillates between good and bad equilibria.
Hence, the distribution of $X_{\hat{\tau}}$ becomes normal and there is no phase transition. The voting rates converge to $(1-p)q+pa/(a+b)\sim(1-p)q+pq^{\frac{r+1}{2}}/(q^{\frac{r+1}{2}}+(1-q)^{\frac{r+1}{2}})$. The first term is the number of votes by independent voters and the  second term is the number of votes by the digital herders. The herders' votes oscillate between good and bad equilibria in (\ref{i3}). As $r$ increases, the stay in the good equilibrium becomes longer. The ratio of stay in the good equilibrium to that in the bad equilibrium is $a/b\sim (\frac{q}{1-q})^{r+1/2}$.

When $r=\infty$, we consider two cases. One is the case where $r$ increases with $t$. Also, in this case, $r$ increases rapidly, that is, the annealing schedule is not adequately slow \cite{ge}. The voters can refer to all historical votes. The hopping rates $a$ and $b$ decrease exponentially as $t$ increases, and  $\bar{v}$ converges to the solutions of the self-consistent equation (\ref{i3}). Hence, above $p_c$, which is the critical $p$, the voting rate could converge to the bad equilibrium. Thus, this is the information cascade transition.

The other case is where $r$ increases slowly, that is, the annealing schedule is adequately slow. At first, we set a finite $r$. After an adequate number of votes, the state frequently oscillates between good and bad equilibria. We increase $r$ after several oscillations of the states. We continue this process until $r$ reaches $\infty$. Thus, we deem the annealing schedule to be adequately slow. (See section 4, where the schedule $r\sim\log t$.)
The ratio of stay in the good equilibrium to the bad equilibrium is $a/b\sim (\frac{q}{1-q})^{r+1/2}\rightarrow \infty$.
It means that the state is always in the good equilibrium, and the voting rate converges only to the good equilibrium $(1-p)q+pa/(a+b)\rightarrow (1-p)q+p$ when  $r$ is large.

\begin{figure}[h]
\includegraphics[width=110mm]{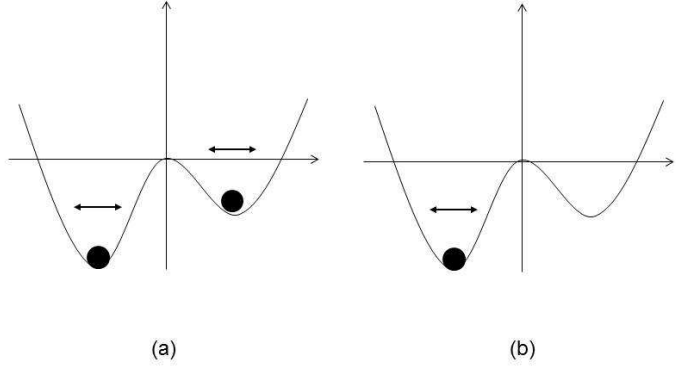}
\caption{Illustration of voting rate convergence  using the analogy of physical potential. The ball stops at the bottom of the potential, corresponding to the convergence of the voting rate. The deeper (shallower) potential corresponds to good (bad) equilibrium. (a) is the case where $r=\infty$. As the annealing schedule is not adequately slow, sometimes, the voting rate converges to the bad equilibrium and there is information cascade. (b) is the case where the annealing schedule is adequate slow. If we increase $r$ slowly, the voting rate converges only to the good equilibrium.}
\label{graph}
\end{figure}

In Figure \ref{graph}, we depict the convergence as $r$ increases. (a) is the case where $r$ increase rapidly, or the annealing schedule is not adequately slow. $\bar{v}$ converges to one of the solutions of the self-consistent equation (\ref{i3}). Hence, the average correct  ratio $E(c_1(t)/t)$ decreases above $p_c$, as $p$ increases. The average correct ratio is the ratio of the number of votes for the candidate $C_1$ to all votes. Further, (b) is the case where $r$ and $p$ increase slowly from finite $r$. In this case, the voting rate converges only to the good equilibrium.

\section{Numerical Simulations}

To confirm the analytic results of section 3, we performed numerical and Monte Carlo (MC) integration of the master equation for the digital herder case. For the Monte Carlo study, we solve the master equation for $10^{6}$ times and calculate the average value of  ratios.

\begin{figure}[htbp]
\begin{center}
\includegraphics[width=10cm]{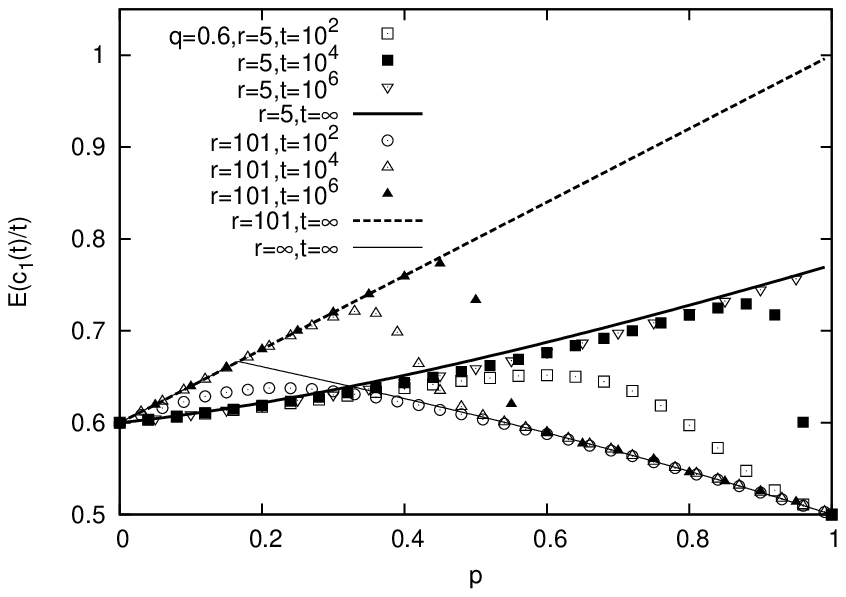}
\end{center}
\caption{$E(c_{1}(t)/t)$ vs $p$ for $r=5,101,\infty$ and
$t=10^{2},10^{4},10^{6},\infty$. The symbols show the results from numerical
studies. The lines represent the theoretical results.
The theoretical results for finite $r$ (\ref{eq:c1_limit})
are plotted with solid ($r=5$) and broken lines ($r=101$).
For $r=\infty$, the result  is given in \cite{Hisakado3} and is plotted with thin solid line.
We set $q=0.6$.
}
\label{r_p_vs_EZ_q06}
\end{figure}

In Figure \ref{r_p_vs_EZ_q06}, we show  the average votes ratio for the correct candidate vs $p$ for $q=0.6$ and $r=5,101$ and $r=\infty$. For the time horizon $t$, we choose $t=10^{2}, 10^{4}$, and $10^{6}$ in order to observe the limit behaviour $t\to \infty$. We also plot the theoretical results for the limit value as
\begin{equation}
\lim_{t\to \infty}E(c_{1}/t)=(1-p)q+p\frac{a}{a+b} \label{eq:c1_limit},
\end{equation}
where $a,b$ are given in (\ref{pi1}).
The thin solid line plots the exact result of the limit value. For $p<p_{c}=1/6$, the herders make the correct choice and it behaves as $(1-p)q+p\cdot 1$. Above $p_{c}$, the probability that all herders choose the wrong candidate becomes finite and it becomes smaller than $(1-p)q+p\cdot 1$.

When $r=5$ and $t=10^{2}$, $E(c_{1}(t)/t)$ slightly increases with $p$ for small $p$. For large $p$, the probability that the system is in the bad equilibrium becomes large and $E(c_{1}(t)/t)$ becomes small. As $p$ increases further, the probability that the system escapes from the bad equilibrium becomes negligibly small and it approaches the limit value for $r=\infty$. As $t$ increases from $10^{2}$ to $10^{4},10^{6}$, the probability that the system is in the bad equilibrium decreases and $E(c_{1}(t)/t)$ becomes larger. As $p$ approaches 1, like in the  $t=10^{2}$ case, it approaches the limit value for $r=\infty$. We also see that $E(c_{1}(t)/t)$ approaches the theoretically estimated limit value (\ref{eq:c1_limit}) as $t$ becomes large. For the limit $t\to \infty$, the probability that herders make the correct choice can be estimated as $a/(a+b)$. For $r=101$, we also see the same feature with $r=5$. Compared with $r=5$, for $r=101$, the peak position of $E(c_{1}(t)/t)$ in $p$-axis becomes smaller and it rapidly approaches the limit value for $r=\infty$. As $r$ becomes large, the mean oscillation time between good and bad equilibria becomes large and the probability that the system escapes from the bad equilibrium becomes small. Despite this, the system finally escapes from the bad equilibrium for the limit $t\to \infty$ and $E(c_{1}(t)/t)$ approaches the theoretically estimated result. The probability that herders make the correct choice $a/(a+b)$ is an increasing function of $r$, and the slope of $E(c_{1}(t)/t)$ vs $p$ becomes larger for large $r$. At $r=100$ and $q=0.6$, $a/(a+b)$ is almost one and $E(c_{1}(t)/t)$ goes to one for  the limit $p\to 1$. For better correct ratio $E(c_{1}(t)/t)$ , $r$ should be large. However, in order to realize high value of $E(c_{1}(t)/t)$, $t$ also should be large. Otherwise, one cannot necessarily realize a high correct ratio. Therefore, there needs to be a trade-off between time and correct ratio.

In order to realise the high value for $E(c_{1}(t)/t)$, $r$ should be large. For large $r$, for the probability in the bad equilibrium to be small, $t$ should be
large and the system should oscillate several times between the two equilibria. As the mean cycle of the oscillation behaves as $1/[(1-p)^{(r+1)/2}(1-q)^{(r+1)/2}]=e^{Cr}$ with $C=-\frac{1}{2}\log (1-p)(1-q)$, a good annealing schedule  $r(t)$ is estimated as
\begin{equation}
r(t)=\log t/C \label{eq:good}.
\end{equation}
The annealing schedule $r(t)$ depends on $t$ logarithmically and
 is very slow.

\begin{figure}[htbp]
\begin{center}
\includegraphics[width=10cm]{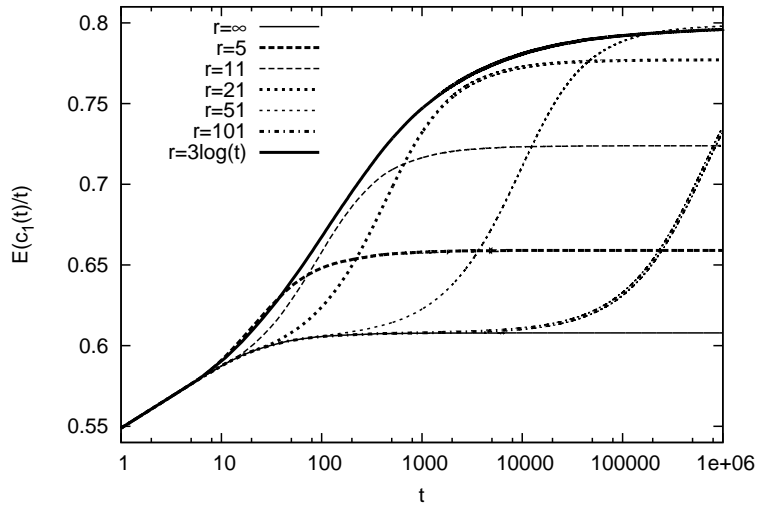}
\end{center}
\caption{Plots of $E(c_{1}(t)/t)$ vs $t$ for $r=5,11,21,51,101,\infty$ and
$r=3\log t$. We set $q=0.6$ and $p=0.5>p_{c}=1/6$.
}
\label{r_t_vs_Z_q06_p05}
\end{figure}

In Figure \ref{r_t_vs_Z_q06_p05}, we plot $E(c_{1}(t)/t)$ vs $t$ for $r=5,11,21,51,101,\infty$ and $r=r(t)$ in eq.(\ref{eq:good}). We set $q=0.6$ and $p=0.5>p_{c}=1/6$. For $r=\infty$, the bad equilibrium is stable and $E(c_{1}(t)/t)$ saturates at low value $\sim 0.6$ for $t\ge 10^{2}$. For $r=5$, the herder's probability of correct choice is low and $c_{1}(t)/t$ soon reaches the maximum value $\sim 0.65$ at about $t=2\times 10^{2}$. For $r=11$, $E(c_{1}(t)/t)$ reaches the maximum value for about $t=10^{3}$. As $r$ increases, the time $t$ needed for the high value of $E(c_{1}(t)/t)$ increases. For $r=51$, $c_{1}(t)/t$ reaches about 0.8 at $t=10^{5}$. If the probability that herders choose the correct candidate $C_1$ is one, $c_{1}(t)/t$ takes $0.8=(1-p)q+1\cdot p$ for $q=0.6$ and $p=0.5$. $r=51$ is large enough to maximise the limit value $\lim_{t\to \infty}c_{1}(t)/t$. If one adopts $r(t)=3\log t$ as the annealing schedule for $r(t)$, $E(c_{1}(t)/t)$ increases smoothly to the maximum value $0.8$ without saturation at some value between $(1-p)q+p\cdot 0.5=0.55$ and the maximum value. If we adopt $r=51$, at $t=10^{6}$, $c_{1}(t)/t$ reaches the maximum value. However, for $t<10^{5}$, $E(c_{1}(t)/t)$ with $r=r(t)$ is higher than that for $r=51$. If we set $r=51$, the probability to escape from the bad equilibrium is small and it is necessary to wait for a long time $t$ for $c_{1}(t)$ to reach the maximum value. In contrast, for $r=r(t)$, $r$ slowly increases and the probability to stay in the bad equilibrium is minimised. Furthermore, unnecessary stay in good equilibrium with medium $E(c_{1}(t)/t)$ is also avoided by increasing $r$ and smoothly increasing $ c_{1}(t)$.


\section{Analog herder and Kirman's ant colony model}
In this section, we consider the case of the analog herder \cite{Hisakado2}. As mentioned earlier, the herders vote for the candidate with the probability that is proportional to the previous votes ratio that can be referred. The voter can see the latest  previous $r$ voters. The transition is
\begin{eqnarray}
c_1(t)&=&k \rightarrow k+1:
 P_{k,t:l,t-r}=(1-p)q+p\frac{k-l}{r}=\frac{q(1-\rho)+\rho (k-l)}{(1-\rho)+\rho r},
\nonumber \\
c_1(t)&=&k   \rightarrow k:
 Q_{k,t:lt-r}=1-P_{k,t:l,t-r}=\frac{(1-q)(1-\rho)+\rho (r-(k-l))}{(1-\rho)+\rho r},
\nonumber \\
\label{pda}
\end{eqnarray}
where $c_1(t-r)=l$.
$P_{k,t:l,t-r}$ and $Q_{k,t:l,t-r}$  are the probabilities of the process.
The voting ratio for $C_1$ is $c_1 (t-r)=l$.
We changed the parameters from $p$ to $\rho$, which is the correlation of $r$th beta binomial model \cite{Hisakado}.
The relation between $p$ and $\rho$ is
\begin{equation}
p=\frac{\rho r}{(1-\rho)+\rho r}.
\end{equation}
Hence, we can map independent voters and herders to beta binomial distribution. As $r$ increases, $1-p$  decreases as $1/r$ and the independent voters' ratio decreases.

Here, we consider the hopping rate among $r+1$ states $\hat{k}=k-l=0,1,\cdots, r$. The dynamic evolution of the process is given by
\begin{eqnarray}
\hat{k} &\rightarrow& \hat{k}+1:
 P_{\hat{k},\hat{k}+1,t}=\frac{r-\hat{k}}{r}\frac{q(1-\rho)+\rho \hat{k}}{(1-\rho)+\rho r},
\nonumber \\
\hat{k} &\rightarrow& \hat{k}-1:
P_{\hat{k},\hat{k}-1,t}=\frac{\hat{k}}{r}\frac{(1-q)(1-\rho)+\rho (r-\hat{k})}{(1-\rho)+\rho r},
\nonumber \\
\hat{k} &\rightarrow& \hat{k}:
P_{\hat{k},\hat{k},t}=1-P_{\hat{k},\hat{k}-1,t}-P_{\hat{k},\hat{k}+1,t}.
\label{pd1}
\end{eqnarray}

This process means that a new vote is added by using the $r$ references, where the oldest vote of the $r$ references exists. If we set
$\epsilon=\frac{1}{2}(1-\rho)/[(1-\rho)+\rho r]$, $1-\delta=(r-1) \rho /[(1-\rho)+\rho r]$, and $q=1/2$, we obtain an equivalent of Kirman's ant colony model \cite{Kir}, \cite{Lux}.
\footnote{Kirman's colony model corresponds to $r\geq 2$ case.}
In our model, there exists a constraint between the parameters,
\begin{equation}
2\epsilon+\frac{r}{r-1}(1-\delta)=1.
\end{equation}

Here, we define $\mu_{r}(\hat{k},t)$ as a distribution function of the state $\hat{k}$ at time $t$.
The number of all states is $r+1$. Using the fact that the process is reversible, we have
\begin{equation}
\frac{\mu_{r}(\hat{k}+1, t)}{\mu_{r}(\hat{k}, t)}=\frac {P_{\hat{k},\hat{k}+1,t}}{ P_{\hat{k}+1,\hat{k},t}}
=\frac{r-\hat{k}}{\hat{k}+1}\frac{q(1-\rho)+\rho\hat{k}}{(1-q)(1-\rho)+(r-\hat{k}-1)\rho}.
\end{equation}
For the limit $t\rightarrow \infty$, we can obtain the equilibrium distribution
\begin{equation}
\mu_{r}(\hat{k}, \infty)={}_{r}C_{\hat{k}}
\frac{\Pi_{j=0}^{\hat{k}-1}(q(1-\rho)+(j\rho))\Pi_{j'=0}^{r-\hat{k}-1}((1-q)(1-\rho)+(j'\rho))}
 {\Pi_{i=0}^{r-1} ((1-\rho)+r\rho)}.
\end{equation}
It is the $(r-1)$th beta binomial distribution that is constructed on the lattice \cite{Hisakado}.
It is the same distribution as that in Kirman's colony model.\footnote{They showed the distribution with one parameter. With the constraint that the sum of the distribution is $1$, it is same as the beta binomial distribution, respectively.} The beta binomial distribution becomes beta distribution for the limit $r
\rightarrow \infty$.

From the central limit theorem, as the distribution of states converges to the beta binomial distribution, the voting  rates for the candidate $C_1$ converges,
\begin{equation}
\lim_{t\rightarrow \infty}\frac{c_1}{t}=\delta_{q},
\end{equation}
where the reference number $r$ is finite and $\delta_q$ is a Dirac measure. Further, when $r\rightarrow \infty$, the distribution of states is the same as the voting rates for the candidate $C_1$, that is, the beta distribution.\footnote{The relations between $\alpha$, $\beta$ of parameters of beta distribution and $q$, $\rho$ are $q=\alpha/(\alpha+\beta)$ and $\rho=1/(\alpha+\beta+1)$.  }

Now, we discuss the case where $r$ increases slowly, that is, annealing schedule is adequately slow. At first, we set a finite $r$. After an adequate number of votes, the distribution of states converges to the $r$th beta binomial distribution. We increase $r$ after the state converges. We continue this process until $r$ reaches $\infty$. We call it the case where the annealing schedule is adequately slow. In this case, the voting  rates  for the  candidate $C_1$ converge to the same point, that is, independent voter's correct ratio $q$, as the finite $r$ case.

In the appendix C, we discuss  the nonlinear extension of this model, Aoki's birth-death processes \cite{Ao}.


\section{Tanh-type herder and Kinetic Ising model}

In this section, we identify the relation between our voting model and the infinite-range kinetic Ising model.

The state of the Ising model is denoted by the vector $\vector{\sigma}=(\sigma_1,\cdots,\sigma_{r+1})$ with $\sigma_j=\pm1$.

The Hamiltonian is defined as
\begin{equation}
H(\vector{\sigma})=-J\sum_{i\neq j}^{r+1}\sigma_i\sigma_j,
\end{equation}
where $J$ is the exchange interaction.

We define $F_j$ as a spin flip operator on $j$th site: $F_j\vector{\sigma}$ is the state in which $j$th spin is flipped from $\vector{\sigma}$ with the other spins fixed. The Markov chain is characterised by a transition probability $w_j(\vector{\sigma})$per unit time from $\vector{\sigma}$ to $F_j\vector{\sigma}$.
Let $p(\vector{\sigma},t_n)$ be  the  probability distribution for finding the spin sate $\vector{\sigma}$ at time $t_n$. Then, the discrete master equation is written as follows:

\begin{equation}
p(\vector{\sigma},t_{n+1})=p(\vector{\sigma},t_{n})-[\sum_j w_j(\vector{\sigma})]p(\vector{\sigma},t_n) \Delta t+[\sum_j w_j(F_j\vector{\sigma})]p(F_j\vector{\sigma},t_n) \Delta t,
\label{master}
\end{equation}
where the second and third terms in RHS are ongoing and incoming probabilities, respectively.

For the master equation (\ref{master}), we consider the sufficient condition that $p(\vector{\sigma},t_n)$ converges to the equilibrium distribution  $\pi(\vector{\sigma})$ as $t_n\rightarrow \infty$ is that the transition probability $w_j(\vector{\sigma})$ satisfies a detail balance condition:
\begin{equation}
w_j(\vector{\sigma})\pi(\vector{\sigma})=w_j(F_j\vector{\sigma})\pi(F_j\vector{\sigma}).
\end{equation}
This condition is known as reversibility. From this condition, the transition probability is given as
\begin{equation}
\frac{w_j(\vector{\sigma})}{F_j w_j(\vector{\sigma})}=\frac{\pi(F_j\vector{\sigma})}{\pi(\vector{\sigma})}=\frac{\exp[-\sigma_j \beta h_j]}{\exp[\sigma_j \beta h_j]},
\end{equation}
where $h_j= J \sum_{i\neq j}^{r+1}\sigma_i$ and an inverse temperature $\beta$, in the units where the Boltzmann constant is $1$.

Here, we set
\begin{equation}
w_j(\vector{\sigma})=\frac{1}{2}(1-\sigma_j\tanh \beta h_j).
\end{equation}

We define the total number of $\sigma_i=1$, where $i\neq j$, is $\hat{c}_1$ and the total number of $\sigma_i=-1$, where $i\neq j$, is $\hat{c}_{-1}$.
Hence, $\hat{c}_{1}+\hat{c}_{-1}=r$. The transition is given by
\begin{eqnarray}
\sigma_j&=&1 \rightarrow -1 :
 w_j(\vector{\sigma})=\frac{1}{2}(1-\tanh  \beta J(\hat{c}_1-\hat{c}_{-1})),
\nonumber \\
\sigma_j&=&-1  \rightarrow 1:
 F_j w_j(\vector{\sigma})=\frac{1}{2}(1+\tanh \beta J(\hat{c}_1-\hat{c}_{-1})).
\label{pda3}
\end{eqnarray}
Note that as $w_j(\vector{\sigma})+F_j w_j(\vector{\sigma})=1$, the transition does not depend on the previous state of $\sigma_j$ and depends on the other spins $\sigma_i$ where $i\neq j$.

In an ordinary case, an  updated spin is chosen randomly. Here, we consider the case where the updated spin is chosen by the rules. The ordering of update is from $\sigma_{1}$  to $\sigma_{r+1}$. After the update of $\sigma_{r+1}$, we update $\sigma_{1}$ and so on. We repeat this process. Hereafter, we define the updated spin $\sigma_j$  after $n$  as  $\sigma_j^{(n)}$. The initial condition is $\sigma_j^{(0)}$, where $j=1,2,\cdots,r$.


Here, we consider a voting model where all herders are $\tanh$-type  herders. The voter can see the latest previous $r$ voters. The transition is given by
\begin{eqnarray}
c_1(t)&=&k \rightarrow k+1:
 P_{k,t:l,t-r}=\frac{1}{2}
[ \tanh \lambda (\frac{k-l}{r}-\frac{1}{2} )+1 ]
\nonumber \\
c_1(t)&=&k   \rightarrow k:
 Q_{k,t:lt-r}=1-P_{k,t:l,t-r},
\label{pda2}
\end{eqnarray}
where $c_1(t-r)=l$ and $\lambda$ is a  parameter. (Please see in \cite{Hisakado4} for details.)
The number of votes for the candidate $C_1$ at time $(t-r)$ is $c_1 (t-r)=l$.

We define a new variable $\Delta_t$ such that
\begin{equation}
\Delta_t=2c_1(t)-t=c_1(t)-c_0(t).
\end{equation}
For convenience, we change the notation from $k$ to $\Delta_t$ as in section 3.
Given $\Delta_t=u$ and $\Delta_{t-r}=s$, we obtain a random walk model:
\begin{eqnarray}
\Delta_t&=&u \rightarrow u+1  :P_{u,t:s,t-r,t}=
\frac{1}{2}(1+\tanh \frac{\lambda (u-s)}{2r}),
\nonumber \\
\Delta_t&=&u \rightarrow u-1  :
Q_{u,t:s,t-r,t}=
\frac{1}{2}(1-\tanh  \frac{\lambda(u-s)}{2r}).
\label{pda4}
\end{eqnarray}

Given that the voter who voted  at  time $(t-r-1)$ chose the candidate $C_0(C_1)$, the probability that another voter at time  $t$ votes to the candidate $C_1 (C_0)$ is
$ P_{u,t:s,t-r,t}(Q_{u,t:s,t-r,t})$.
Hence, (\ref{pda4}) means that the hopping rate is $P_{u,t:s,t-r,t}$  and $ Q_{u,t:s,t-r,t}$.
This is equivalent to (\ref{pda3}), where $\beta J=\lambda/2r$, $\hat{c}_1-\hat{c}_{-1}=u-s$.
We use the relations $\hat{c}_{1}=k-l$, $\hat{c}_{-1}=r-(k-l)$, $u=2k-t$, and $s=2l-(t-r)$.

If  we consider the row of spins $ \sigma^{(0)}_1\cdots \sigma^{(0)}_{r+1}\sigma^{(1)}_1\cdots\sigma^{(n)}_{r+1}\sigma^{(n+1)}_1\cdots$  where $n=0,1,2 \cdots$ as the  voting row. Hence, the voting model is equivalent to the infinite-range kinetic Ising model.
This is why we can find the same mean field approximation   equation  as  that in the Ising model in the  limit $r\rightarrow \infty$ case \cite{Hisakado4}.
Thus, for the finite $r$ case, we confirm that there is no phase transition \cite{Hisakado3} and that it is not appropriate to use the mean field approximation as well.

When $r\rightarrow \infty$, the kinetic Ising model is the voting model when the annealing schedule is adequately slow, and the state is always in the good equilibrium.
Hence, the Ising model with external field has no phase transition, as discussed in sections 3 and 4.
In fact, we can confirm as $r$ increases, the maximum point of $E(c_1(t)/t)$ approaches the maximum point of $r=3\log (t)$, the slow annealing, in Fig. 4.
On the other hand, when the  annealing schedule is not adequately slow, the voting model has the phase transition.
It means that the voting rate could converge to the bad equilibrium.
The phase transition is observed when the size of $r$  is infinite.

In addition, the voting model becomes  the Potts model, if there are several candidates above two.(See Appendix D.)
This is a simple extension of this section.
While in this section, we have discussed the Ising model without external fields, in Appendix E, we discuss how prior distribution and independent voters correspond to the outer fields of the Ising model.

\section{Concluding Remarks}

In this study, we investigated a voting model that involves collective herding behaviours. We investigated the case where voters can obtain the information from previous $r$ voters. We observed the states of reference votes and considered three cases of voters:

(i) digital herders and  independent voters,

(ii) analog herders and independent voters, and

(iii) $\tanh$-type herders.

In the previous paper \cite{Hisakado3}, we investigated the case where there were only digital herders and independent voters.
By using the mean field approximations, we concluded that if $r\geq 3$,
there occurs a phase transition from the one-peak phase to the two-peaks phase.
In this paper, through numerical simulations and analysis, we show that for finite $r$, no phase transitions occur and that there is only one peak phase.

From the viewpoint of annealing, if the annealing schedule is adequately slow, that is, $r\sim \log t$, the voting rate converges to the good equilibrium only. The correct ratio $E(c_1(t)/t)$ increases as $r$ increases. On the other hand, if the annealing schedule is not adequately slow, that is, $r\sim t$, the phase transitions occur.
Further, above $p_c$, there are good and bad equilibria.
The voting rates converge to one of the two equilibria, but we cannot estimate which of these equilibria is selected.
The correct ratio decreases when phase transition exists.
Therefore, there exists a trade-off between time and the correct ratio.

In case (ii), we considered analog herders and independent voters.
We show that the states of reference votes in this model are equivalent to Kirman's ant colony model.
In this case, the states follow  beta binomial distribution and  for  large $r$, this becomes beta distribution.

In case, (iii)  we show that  the model is equivalent to the finite-size kinetic Ising model.
This model follows a simple herding experiment
if the voters are rational \cite{Bikhchandani}, \cite{An}.
When $r\rightarrow \infty$, the voting model behaves as  the infinite-range   kinetic Ising model when the annealing schedule is adequately slow.
Hence, the Ising model with external field has no phase transition.
On the other hand, when the annealing schedule is  not adequately slow,
the voting model with external fields  has the phase transition.
While there are many studies in which the herding behaviour has been analysed by the Ising model \cite{nuno},
only the current study explains this behaviour by using the kinetic Ising model.

In this paper, we investigated the model on the 1D extended lattice.
Non-regular topologies have non-trivial effects on the dynamics of the voter model \cite{Cas}\cite{Mor}.
Determining how these network effects change $pc$and $pvc$ is one of our future problems.

 \appendix\section*{Acknowledgement}
This work was supported by Grant-in-Aid for Challenging
Exploratory Research 25610109.


\appendix
\def\thesection{Appendix \Alph{section}}
\section{Derivation of stochastic differential equation}

We use  $\delta X_\tau=X_{\tau+\epsilon}-X_\tau$ and $\zeta_\tau$, a standard i.i.d. Gaussian sequence; our objective is to identify the drift $f_\tau$ and  the variance $g^2_\tau$  such that
\begin{equation}
\delta X_\tau=f_\tau(X_\tau)\epsilon+\sqrt{\epsilon}g_\tau(X_\tau)\zeta_{\tau+\epsilon}.
\end{equation}
Given $X_\tau=x$, using the transition probabilities of $\Delta_n$, we get
\begin{eqnarray}
\textrm{E}(\delta X_\tau)&=&\epsilon \textrm{E}(\Delta_{[\tau/\epsilon]+1}-\Delta_{[\tau/\epsilon]})=\epsilon(2p_{[\frac{l/\epsilon+\tau/\epsilon}{2}],\tau/\epsilon}-1)
\nonumber \\
&=&
\epsilon[(1-p)(2q-1)-p+2p\frac{(2n+1)!}{(n!)^2}\int_0^{\frac{1}{2}+\frac{X_\tau-X_{\tau-\hat{r}}}{2\hat{r}}}x^n(1-x)^ndx ].
\nonumber \\
\end{eqnarray}
Then, the drift term is $f_\tau(x)=(1-p)(2q-1)+p +2p\frac{(2n+1)!}{(n!)^2}\int_0^{\frac{1}{2}+\frac{X_\tau-X_{\tau-\hat{r}}}{2\hat{r}}}x^n(1-x)^ndx $.
Moreover,
\begin{equation}
\sigma^2(\delta X_\tau)=\epsilon^2
[
1^2p_{[\frac{l/\epsilon+\tau/\epsilon}{2}],\tau/\epsilon}
+(-1)^2(1-p_{[\frac{l/\epsilon+\tau/\epsilon}{2}],\tau/\epsilon})]
=\epsilon^2,
\end{equation}
  such  that
$g_{\epsilon,\tau}(x)=\sqrt{\epsilon}.$
We can obtain $X_\tau$ such that it obeys a diffusion equation with small additive noise:
\begin{equation}
\textrm{d}X_\tau=[(1-p)(2q-1)-p+ 2p\frac{(2n+1)!}{(n!)^2}\int_0^{\frac{1}{2}+\frac{X_\tau-X_{\tau-\hat{r}}}{2\hat{r}}}x^n(1-x)^ndx ]\textrm{d}\tau+\sqrt{\epsilon}.
\label{ito}
\end{equation}

\section{Analysis of relaxation time}

In order to see the relation between the oscillation time and $r$, we estimate the relaxation time or integrated correlation time $\tau$ for the digital herder case. We denote the $t$th vote as $X(t)$, which takes $1(0)$ if the voter opts for the candidate $C_1(C_0)$.
We estimate the auto-correlation function $d(t)$ as the covariance between $X(1)$ and $X(t+1)$.
That is,
\[
d(t)\equiv \mbox{Cov}(X(1),X(t+1)).
\]
The integrated correlation time $\tau(t)$ for time horizon $t$ is defined as
\[
\tau(t)\equiv \sum_{s=0}^{t-1}d(s)/d(0).
\]
For $r=1$, one can show that $d(t)$ and $\tau(t)$ behaves as
\begin{eqnarray}
d(t)&=&(q(1-q)+(q-\frac{1}{2})^{2}(2p-p^{2}))\cdot p^{t-1} \nonumber \\
\tau(t)&=&\frac{1-p^{t}}{1-p}. \nonumber
\end{eqnarray}
We are interested in the extrapolated value
$\lim_{t\to \infty}\tau(t)$ to the limit $t\to \infty$,
which we denote as $\tau$.
For $r=1$, we have
\begin{equation}
\tau=\frac{1}{1-p}.
\end{equation}

\begin{figure}[h]
\begin{center}
\begin{tabular}{c}
\begin{minipage}{0.5\hsize}
\begin{center}
\includegraphics[clip, width=6.5cm]{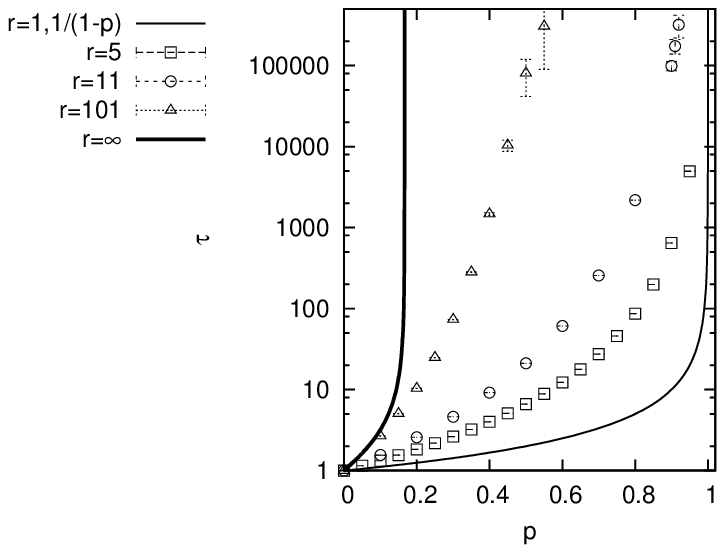}
\hspace{1.6cm} (a)
\end{center}
\end{minipage}
\begin{minipage}{0.5\hsize}
\begin{center}
\includegraphics[clip, width=6.5cm]{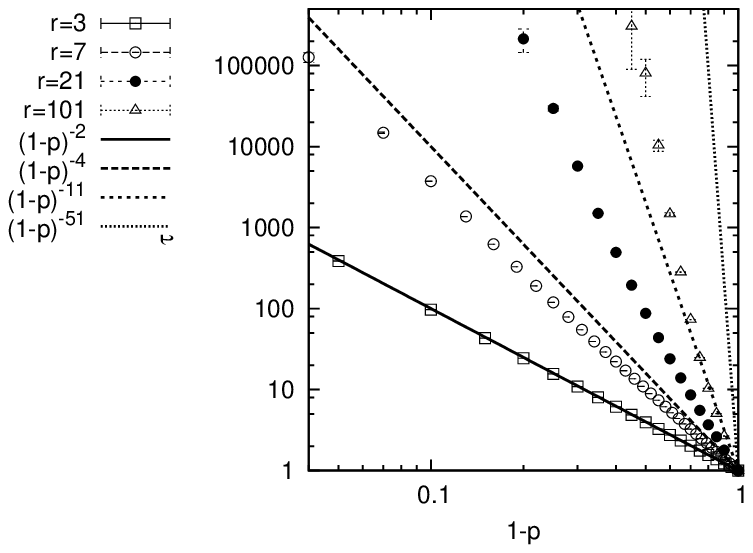}
\hspace{1.6cm} (b)
\end{center}
\end{minipage}
 \end{tabular}
\caption{ Plot of $\tau$ vs $p$ (a) and $1-p$ (b).
$\tau$ is the limit value $\tau=\lim_{t\to \infty}\tau(t)$.
The left figure is semi-logarithmic plot
for $r=1$ (thin solid), 5 (box), 21 (circle),
101 (triangle) and $\infty$ (thick solid).
The right figure is
double-logarithmic plot of $\tau$ vs $1-p$ for $r=3$ (box),$7$ (circle), $21$ (filled circle), and $101$(triangle). We also plot $1/(1-p)^{(r+1)/2}$.
We estimate the error bars as the absolute value of the
discrepancies of the extrapolated values for
$T=4\times 10^{3}$ and $2\times 10^{3}$.
We set $q=0.6$.}
\label{syml}
\end{center}
\end{figure}

In Figure \ref{syml}, (a) plots $\tau$ vs $p$ for $r=1,5,11,101$ and $\infty$. In general, $\tau$ is a monotonic increasing function of $p$ and $r$.
For $r=\infty$, as the system shows the phase transition at $p_{c}=1/6$, $\tau$ diverges at $p_{c}$. For $r=1$, $\tau$ behaves as $1/(1-p)$
and it diverges only at $p=1$. It reflects that the system for $r=1$ does not show any phase transition for $p<1$.
The mean cycle of the oscillation, which is proportional to $\tau$, is estimated as $1/a+1/b\simeq 1/[(1-p)^{(r+1)/2}(1-q)^{(r+1)/2}]$.
$\tau$ diverges as $(1-p)^{-(r+1)/2}$ in the limit $p\to 1$ for finite $r$. As $r$ increases, $\tau$ rapidly increases for the same $p$. For $r=101$ and $p=0.5$, $\tau$ takes the value of the  order of magnitude  $10^{5}$. In order to escape from the bad equilibrium, $t$ should be as large as the order of  $10^{6}$.
In the case, $E(c_{1}(t)/t)$ takes $(1-p)q+p\cdot 1$ (Figure \ref{r_p_vs_EZ_q06}).

In Figure \ref{syml}, (b) shows the double-logarithmic plot of $\tau$ vs $1-p$ for $r=3,7$ and $101$.
For $r=3$, $\tau$ diverges as $(1-p)^{-2}$, which is consistent with $(1-p)^{-(r+1)/2}$ for $r=3$.
For large $r$, it is difficult to estimate $\tau$ in the limit $1-p\to 0$, one can see that the absolute value of the
slope rapidly increases with the increase of $r$.

At last, we estimate the relaxation time $\tau$ for $r=\infty$ when  the analog herder case. The covariance $d(t)$ between $X(1)$ and $X(t+1)$ is calculated as
\[
d(t)=(q(1-q)+(q-\frac{1}{2})^{2}(2p-p^{2}))
\prod_{s=0}^{t-1}\left(\frac{s+p}{s+1}\right).
\]
As $d(t)$ behaves as $t^{p-1}$ for large $t$, the relaxation time $\tau(t)$ for time horizon $t$ is estimated as
\[
\tau(t) \simeq \int_{0}^{t}t^{p-1}dt=\frac{1}{p} t^{p}.
\]
$\tau(t)$ diverges as $t^{p}$ for $p>0$ in the limit $t\to \infty$. This means that the oscillation time is infinite for $r=\infty$. For $r<\infty$, one can show that $d(t)$ decays exponentially for large $t$ and $\tau(t)$ becomes finite in the limit $t\to \infty$ for $p<1$. The result is consistent with the result for the digital herder model, that $\tau$ is finite for $r<\infty$ and $p<1$.

\section{Aoki's  Birth--Death Processes}

In this section, we discuss Birth--Death Processes \cite{Ao}.
The model is a nonlinear extension of Kirman's ant colony model \cite{Kir}.

We consider  the voting model (\ref{pda2}),
where we consider the hopping rate among $r+1$ states $\hat{k}=k-l=0,1,\cdots, r$.
The dynamic evolution of  the process is  given by
\begin{eqnarray}
\hat{k} &\rightarrow& \hat{k}+1:
 P_{\hat{k},\hat{k}+1,t}=\frac{1}{2}\frac{r-\hat{k}}{r}[1+\tanh \lambda(\frac{\hat{k}}{r}-\frac{1}{2})],
\nonumber \\
\hat{k} &\rightarrow& \hat{k}-1:
P_{\hat{k},\hat{k}-1,t}=\frac{1}{2}\frac{\hat{k}}{r}[1-\tanh \lambda(\frac{\hat{k}}{r}-\frac{1}{2})],
\nonumber \\
\hat{k} &\rightarrow& \hat{k}:
P_{\hat{k},\hat{k},t}=1-P_{\hat{k},\hat{k}-1,t}-P_{\hat{k},\hat{k}+1,t}.
\label{aoki}
\end{eqnarray}
This is the case
\begin{eqnarray}
\eta_1(\hat{k})&=&\frac{1}{2}[1+\tanh \lambda(\frac{\hat{k}}{r}-\frac{1}{2})],
\nonumber \\
\eta_2(\hat{k})&=&\frac{1}{2}[1-\tanh \lambda(\frac{\hat{k}}{r}-\frac{1}{2})],
\end{eqnarray}
in \cite{Ao}.
From the discussion in section 6, we conclude that this is the Ising model.

\section{Potts model and  Information cascade model}
Now, we study  Bayes' model for information cascade model \cite{Bikhchandani}. The original experiment is the estimation of the correct pot by the voting.
The difference between our voting model and the pot model is as follows. The voter casts his/her vote by using original information as well as information on the previous votes. The original information is given to each voter in the form of choice between candidates $C_1$ and $C_0$. We assume that the correct candidate is the candidate $C_1$. Note that in this case, we do not distinguish between the herders and independent voters. The voter estimates the posterior distribution
where $C_i$ where $i=0,1$ is the correct candidate by using Bayes' theorem.

In this section, we consider the extended model of this model. While originally in this study, the voters choose between two candidates,
here, we consider that the voters choose among $\omega$ candidates $C_i$ where $i=1.2.\cdots, \omega$.
We assume that the correct candidate is the candidate $C_1$.

Voters estimate the probability that $C_i$, where $i=1,2,\cdots,\omega$, is the correct candidate by using original information and the information on $(r-1)$ previous votes.
The voters do not distinguish between original information and reference votes. Hence, the voters have the $r$ votes as the information.
$Pr(C_i)$ is the probability that the voter estimates that $C_i$ is the correct candidate.
We set the prior distribution as $Pr(C_i)=1/\omega$.
Here, we assume that the voters estimate the percentage of correct answers as $\hat{q}$.
The sum of previous voters' votes and the original information for the candidate $C_i$ is $k_i$.
The total votes are $r$, $r=\sum_{i=1}^{\omega}k_i$.
The posterior distribution is
\begin{equation}
Pr(C_i| c_1^r=k_1, \cdots, c_{\omega}^r=k_\omega)=\frac{1}{\omega}\frac{r!}{(k_1)!\cdots(k_\omega)!}\hat{q}^{k_i}(1-\hat{q})^{r-k_i},
\end{equation}
where $c_i^r=k_i$ indicates that the number of votes with original information for $C_i$ is $k_i$ before the individual votes.
Then, we obtain
\begin{equation}
\frac{Pr(C_{j_1}|c_1^r=k_1, \cdots, c_{\omega}^r=k_{\omega})}{Pr(C_{j-2}|c_1^r=k_1, \cdots, c_{\omega}^r=k_{\omega})}=(\frac{\hat{q}}{1-\hat{q}})^{k_{j_1}-k_{j_2}}=\frac{e^{{\lambda}{k_{j_1}/r}}}{e^{{\lambda}{k_{j_2}/r}}},
\label{pda6}
\end{equation}
where $\lambda= r \log\frac{\hat{q}}{1-\hat{q}}$.

The state of the Potts model is denoted by the vector
$\vector{\sigma}=(\sigma_1,\cdots,\sigma_{r+1})$
with $\sigma_j=1,2,\cdots, \omega$.
The Hamiltonian is defined as
\begin{equation}
H(\vector{\sigma})=-J\sum_{i\neq j}^{r+1}\delta(\sigma_i,\sigma_j),
\end{equation}
where $J$ is the exchange interaction.

Let $p(\vector{\sigma},t_n)$ be a probability distribution for finding the spin sate $\vector{\sigma}(\sigma_j=j_1)$ at time $t_n$.
We define $F_j(j_1,j_2)$ as a spin flip operator on $j$th site to be the state that $j$th
spin is flipped from $j_1$ to $j_2$  with the other spins  fixed, where the $j$th spin of $\vector{\sigma}$ is $j_2$:$F_j(j_1,j_2)\vector{\sigma}(\sigma_j=j_1)=\vector{\sigma}(\sigma_j=j_2)$.

We define a transition probability $w_j(\vector{\sigma})$per unit time from $\vector{\sigma}$
to $F_j(j_1,j_2)\vector{\sigma}$.
$w_j(\vector{\sigma})$ is the probability of  that  $\sigma_j$  hops from $j_1$ to $j_2$.
$w_j(F_j(j1,j2)\vector{\sigma})$ is the probability of that
$\sigma_j$ hops from $j_2$ to $j_1$.

We consider that $p(\vector{\sigma},t_n)$ converges to the equilibrium distribution $\pi(\vector{\sigma})$
as $t_n\rightarrow \infty$  is when the transition probability $w_j(\vector{\sigma})$ satisfies a detail balance condition:
\begin{equation}
w_j(\vector{\sigma})\pi(\vector{\sigma})=w_j(F_j(j_1,j_2)\vector{\sigma})\pi(F_j(j_1,j_2)\vector{\sigma}).
\end{equation}

From this condition, the transition probability is given by
\begin{equation}
\frac{w_j(\vector{\sigma})}{F_j(j_1,j_2) w_j(\vector{\sigma})}=\frac{\pi(F_j(j_1,j_2)\vector{\sigma})}{\pi(\vector{\sigma})}=\frac{\exp[\beta (h_j(j_2)-h_j(j_1))]}{\exp[\beta (h_j(j_1)-h_j(j_2))]}
=e^{2\beta(h_j(j_2)-h_j(j_1))},
\label{pda5}
\end{equation}
where $h_j(\sigma j)= J \sum_{i\neq j}^{r+1}\delta(\sigma_j, \sigma_i)$ and an  inverse temperature $\beta$, in the units where Boltzmann constant is $1$.
$h_j(\sigma j)$ is the number of $\sigma=\sigma_j$ without  $\sigma_j$.

From (\ref{pda6}) and (\ref{pda5}), if we set $\lambda/2r=\beta J$, $k_{j_1}=h_j(j_1)$ and $k_{j_2}=h_j(j_2)$, we identify that these processes are the same.

As $r$ increases, the voting ratio converges to one of the equilibria.
The large $r$ limit in $\omega=2$ case represents the original information cascade experiment.
The voting models and information cascade experiments  with $\omega$ candidates are equivalent to the Potts model.

In \cite{nuno}, the real vote  data are analysed using Potts model, and it is shown that the voting behaviour adheres to the model from the macro perspective using the real voting data.
On the other hand, we have shown the voters behave as in the Potts model from a micro perspective.

\section{Outer field of Ising model}

In this section, we discuss the prior distribution and independent voters play the role of outer fields in Ising model.
We consider the case where two candidates $C_1$ and $C_0$.
We assume the correct candidate to be the candidate $C_1$.

Voters estimate the probability that $C_i$, where $i=0,1$, is the correct candidate by using the $(r-1)$  previous votes and original information.
The voters do not distinguish between  original information and reference votes.
Hence, the voters have the $r$ votes as the information.
$Pr(C_i)$ is the probability that the voter estimates that $C_i$ is the correct candidate. The sum of previous voters' votes and the original information for the candidate $C_1$ is $c_1^r =k$.
Total number of votes for the candidate $C_0$ and $C_1$ is $r$.
We set the prior distribution as
$Pr(C_1)=\hat{p}$.
Here, we assume that the voters estimate the percentage of correct answers as $\hat{q}$.

We obtain, as in the previous section,
\begin{equation}
\frac{Pr(C_1|c_1^r =k)}{Pr(C_0|c_1^r =k)}=(\frac{\hat{p}}{1-\hat{p}})(\frac{\hat{q}}{1-\hat{q}})^{2k-t}=e^{{2\lambda}{(k-\frac{r}{2})/r}+2h},
\end{equation}
where $\lambda= t \log\frac{\hat{q}}{1-\hat{q}}$ and $h=\frac{1}{2}\log \frac{\hat{p}}{1-\hat{p}}$.
We obtain the correspondence,
\begin{equation}
\frac{\hat{h}}{J}=\frac{h}{\beta J}=\frac{2r}{\lambda}h,
\end{equation}
where $\beta J=\lambda/2r$.
$\hat{h}=h/\beta$ corresponds to the outer fields in the Ising model.
In \cite{Mori6}, we discussed the analog herder case with prior distribution as the outer field when there are three candidates or more. Thus, if herder is the  tanh-type, the model is equivalent to  the Potts model with outer field. (See the Appendix D.)

\end{document}